**Core Elements in the Process of Citing Publications:**

**A Conceptual Overview of the Literature**

Iman Tahamtan[1] & Lutz Bornmann[2]

1. Corresponding Author. School of Information Sciences, College of Communication and Information, University of Tennessee, Knoxville, TN, USA. Email: tahamtan@vols.utk.edu

2. Administrative Headquarters of the Max Planck Society, Division for Science and Innovation Studies, Hofgartenstr. 8, 80539 Munich, Germany. Email: bornmann@gv.mpg.de

**Abstract**

This study provides a conceptual overview of the literature dealing with the process of citing documents (focusing on the literature from the recent decade). It presents theories, which have been proposed for explaining the citation process, and studies having empirically analyzed this process. The overview is referred to as conceptual, because it is structured based on core elements in the citation process: the context of the *cited document*, processes *from selection to citation of documents*, and the context of the *citing document*. The core elements are presented in a schematic representation. The overview can be used to find answers on basic questions about the practice of citing documents. Besides understanding of the process of citing, it delivers basic information for the proper application of citations in research evaluation.

**Key words**

# 1   Introduction

Citations in scholarly publications are used in very different contexts of research evaluation, which focuses on measurements of research performance, scholarly quality, influence, or impact (Moed, 2017, Moed, 2005). Citations are used to compare the performance of universities worldwide (e.g. Waltman et al., 2012), to analyze the impact of documents published by single scientists (e.g. Bornmann and Marx, 2014), to reveal citation classics and landmark papers in a field (see www.crexplorer.net), and to study collaborations between institutions worldwide (Bornmann et al., 2016). The use of citations in these and similar contexts has its roots in citation theories, which have been proposed in the past (see an overview in Cronin, 1984, Nicolaisen, 2007, Davis, 2009, Moed, 2005). The first and most prominent theory is the normative citation theory (proposed by Merton) where documents are cited if they have influenced the author of the citing document (Merton, 1973). Merton (1973) provides a theoretical basis for scientometrics, in which citations indicate peer recognition through mechanisms such as awards. His view on citations serves as a basis for the use of citations in performance measurements: more citations mean more recognition. According to Merton (1973), scientists are motivated to cite their peers, by their belief in the justice of giving credit, and the hope of increasing the likelihood of receiving credit through peer recognition.

The theory has been heavily criticized because it explains only a subset of citation decisions (or nothing at all). Many other factors besides cognitive influence and peer recognition have been identified in the past. These other factors are mostly regarded as confirmation of the social-constructivist theory of citing. In social-constructivist theory, citations are seen as rhetorical devices which are not related to the theory of Merton (1973). The social-constructivist theory questions the validity of the normative assumption of the use of citations as reward. Citations are seen as complex processes which cannot be captured by cognitive influence alone. A typical and highly cited paper in this context is Gilbert (1977) who regards citations as tools for persuasion. According to Gilbert (1977), an author selects documents for citing which were published by reputable authors in the field. Thus, the cited documents have not been selected because of their content, but in order to influence the reader as to the claims of the citing author.

Nicolaisen (2004) has critiqued the normative and constructivist theory of citing and proposed another citation theory (also see Nicolaisen and Frandsen, 2007, Nicolaisen, 2007) which is rooted in the so-called handicap principle, developed by Zahavi and Zahavi (1999). Nicolaisen (2004) claims that (human) citation behavior can be explained by theories of honesty and deception in animal communication. He argues that references may be seen as threat signals similar to those in nature such as approaching a rival. The potential cost of dishonest referencing, specifically when the citing references are made public, would make authors reconsider their deceiving behavior. A skilled author detects the false reference and then know where to criticize. Authors would usually not make the risk of losing their reputation by using weak or dishonest references. Nicolaisen (2007, p. 629) suggests that *"the handicap principle ensures that citing authors honestly credit their inspirations and sources to a tolerable degree-enough to save the scientific communication system from collapsing"*. He also notes that the level of honesty and deceit varies across scientific communities. A higher level of deceit might be seen in young and immature fields with a less attack by skilled authors (Nicolaisen, 2004). Nicolaisen's theory has not been formally tested. Yet, it has attracted attention from other researchers, such as Small (2010). Small (2010 p. 192) discusses Nicolaisen's theory, and agrees that a citation theory based on evolutionary theory is "a fruitful topic for further research".

The current study is intended to synthesize the empirical literature on citations, which is mostly rooted in the normative or social-constructivist theories. The study has not planned to be a complete review of the extensive literature, but focusses on historical landmark papers and the literature published in recent years (since 2008). Similar overviews of studies dealing with factors influencing citations and important elements in the citation process have been published a decade ago (Bornmann and Daniel, 2008b, Nicolaisen, 2007). The current overview can be used in the evaluative practice to know the various elements, which are relevant in the citation process. It helps to understand and interpret the results of citation analyses in the context of research evaluations.

This study is designed as a conceptual overview which is structured according to three core elements in the process of citation: the context of the *cited document*, processes *from selection to citation of documents*, and the context of the *citing document*. The core elements are presented in a schematic representation. Many empirical studies focusing on the process of citing have been published to identify factors influencing the number of citations. For example, it has been shown in several studies that the importance of the journal in a field has an influence on the citation impact of the papers published in the journal (Tahamtan et al., 2016). Another group of studies (mostly from recent years) have investigated the context of citations (see, e.g., Boyack et al., 2013, Zhao and Strotmann, 2014, Jha et al., 2016). In these studies, the words and sentences around citations are analyzed to get to know information about characteristics of the cited work, reasons to cite, and decision rules of the citing authors (Halevi and Moed, 2013). In a very recent study, for example, Small et al. (2017) investigated the context of a set of citations to find words which characterize the cited research as discoveries in science. An overview of the different approaches for undertaking context-based citation analysis can be found in Ding et al. (2014).

## 2   Methods: search for the literature

The search for the literature was conducted in 2017. We systematically searched publications of all document types (journal articles, reviews, collected works, monographs, etc.). In a first step, we used the tables of contents of journals in the area of information science, including *Journal of Documentation*, *Scientometrics, Journal of the Association for Information Science and Technology,* and *Journal of Informetric.* We identified relevant papers by reading their titles and abstracts in the table of contents of these journals. In a second step, we analyzed reference lists provided by former reviews of the literature on the citation process to locate relevant studies, including Bornmann and Daniel (2008b), Nicolaisen (2007), Wang and White (1999), Tahamtan et al. (2016), Wang and Soergel (1998), and Erikson and Erlandson (2014). The received literature from these two search steps are used as a basis for further searches in literature databases: (i) we prepared a bibliogram (White, 2005) to obtain keywords for the studies located in the two search steps. The bibliogram ranks the words included in the titles and abstracts of the publications located. Frequent words at the top of the list (e.g., *citation analysis, theory of citation,* and *citation theory*) were used for searches in Web of Science (WoS, Clarivate Analytics), Scopus, and PubMed, as well as Google Scholar. (ii) We identified all citing publications for a series of highly cited papers (found in the two search steps).

## 3   Core elements in the process of citation

In this conceptual overview of the literature, we used three core elements for structuring the results of the found empirical studies: *cited document*, *from selection to citation*, and *citing document* (see Figure 1). Thus, we distinguish between the document which is cited (referred to as the cited document) and the document which cites (referred to as the citing document). For both document types, the empirical studies focus on the relevance of the features pertinent to the document, author, or the journal for the citation process. The connection between the cited and citing document is established by the process from selection to citation. This process is characterized by specific reasons to cite and decision rules of selecting documents for citing. In the following sections, various empirical studies are presented which we have assigned to these core elements. Thus, the schematic representation in Figure 1 serves as a conceptual condensation of the extensive empirical literature on citations.

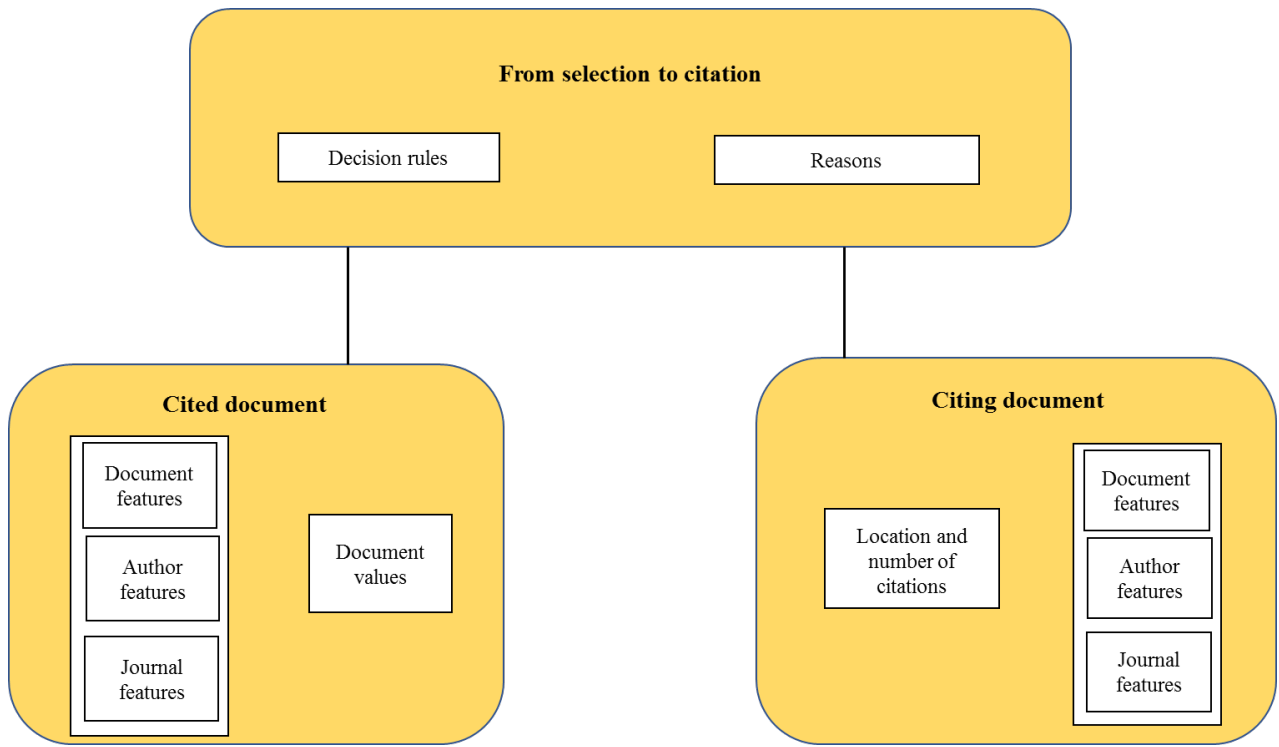

Figure 1. Core elements in the process of citing

## 3.1   Cited document

### 3.1.1   Document features

Generally, citing authors assess selected documents based on several criteria, which are described in some previous studies (Wang and White, 1999, Tahamtan et al., 2016, Wang and Soergel, 1998). Title, abstract, and keywords are among the elements according to which the value of a document is primarily assessed. Thus, citing authors select or reject a document based on its title, abstract, and/or keywords. These elements are the basis for the assessment of many other features of the document, such as its topicality (Wang and Soergel, 1998). If the citing authors perceive these elements as relevant, the document's full text has a chance of being cited.

Several studies have indicated that some characteristics of a document's title and abstract might be less or more attractive for different readers. These features include: diversity and number of keywords in title or abstract (Falagas et al., 2013, Annalingam et al., 2014), length of title and abstract (Stremersch et al., 2015), the presence of certain words in the document's abstract (Ibáñez et al., 2009), type of abstract (structured or unstructured) (Lokker et al., 2008), the presence of punctuation marks, such as hyphens, commas, colons, and brackets in the title (Buter and van Raan, 2011), the names of certain countries in the title (Jacques and Sebire, 2010), the type of title (compound title, question title, or descriptive title) (Subotic and Mukherjee, 2014), and report of the study design in the title (Antoniou et al., 2015). For instance, documents with shorter titles might attract more attention than documents with longer titles (Ayres and Vars, 2000). An important document feature is topicality, which can be defined as the level of the document's relevance to the topic of the citing document (Wang and Soergel, 1998). In general, hot topics have a higher topicality for citing authors, and in turn attract more attention and citations from the scientific community. Researchers usually prefer to write about hot topics in their areas and thus cite the corresponding literature (Fu and Aliferis, 2010, Gallivan, 2012).

Another feature according to which the document value is assessed is its perceived quality. If a document is considered to be of poor quality, it may be less likely to be read and cited even if it is physically and cognitively accessible (Liu, 1997). There is no consensus on the definition of quality and how it should be measured (Hug et al., 2014). It is a more or less subjective construct with no generalizable definition, and is assessed differently in disciplines. Researchers have differing attitudes toward quality, so that some consider accuracy and importance of research as higher quality, while for others it is creativity and novelty (Tahamtan et al., 2016). Citing authors frequently assess quality based on the prestige of a document's journal, authors, institutions, and its citation status (whether it has been already cited by others) (Wang and Soergel, 1998).

Accessibility of a document is also a major criterion for assessing a document's value. This feature refers to the document's availability and observability through information resources, such as the library collection, interlibrary loan service, online sources, or personal contacts. Accessibility and visibility are mentioned in a large number of studies as impacting on citation decisions (Henneken et al., 2006, Rees et al., 2012, Yue and Wilson, 2004). Citing authors might not consider availability when searching and selecting the document, but the document might not have the opportunity of being read if it is hard to obtain (Wang and Soergel, 1998).

Other document features that are considered by citing authors include orientation and level (Wang and Soergel, 1998). Orientation refers to the document type and study design. For instance, whether the document is theoretically or empirically oriented, or whether it is a methodology paper or review. Level refers to the audiences that the document is intended for, including academia or the public. Citing authors most often assess orientation and level by title, abstract, table of content, and/or journal (Wang and Soergel, 1998). Authors tend to read and cite specific types of documents, such as reviews or methodological studies. The former provides readers with comprehensive information and the latter introduces new scientific tools (Padial et al., 2010).

In a time of broad access to citation indexes, authors tend to cite documents that receive their initial citations very soon after publication. When a document is cited, researchers observe that, and this increases their interest in the document and the likelihood of citing (accumulative advantage). Frandsen and Nicolaisen (2017) compared 67,578 pairs of studies on the same healthcare topic, with the same publication age (1-15 years) and showed that documents selected for citation, have on average received about three times as many citations as unselected documents – slightly more for younger studies than older. However, if a new publication captures the attention previously paid to an older publication, the cycle may be broken and the new document might be more attractive for citing (Small, 2004).

Further document features, which are reviewed by Tahamtan et al. (2016), include, e.g. characteristics of a document's results, discussion, references, and the document's length. For instance, it has been revealed that a positive correlation exists between citation counts and the number of cited references (Falagas et al., 2013) that is documents with more cited references would achieve more citations (Vieira and Gomes, 2010). Specifically, review papers usually have long lists of references and are heavily cited which inflates, e.g. the journal impact factor (JIF) – a well-known journal metric (Seglen, 1997). For this reason, recent studies have attempted to construct journal impact metrics that try to correct for this. Perhaps most notably is the so-called Reference Return Ratio (3R) introduced by Nicolaisen and Frandsen (2008). "By taking the number and age of references into account, the 3R avoids some of the problems facing the traditional impact measures" (Nicolaisen and Frandsen, 2008 p. 128-129).

Webster et al. (2009, p. 356) indicate that in evolutionary psychology, "reference counts explained 19% of the variance in citation counts". One possible explanation for such relationship is that "the more people you cite in your paper, the more people are likely to cite your paper (the paper they were cited in) in the future" (Webster et al., 2009, p. 349). In addition, the citation performance of the cited references can be correlated with the citation counts (Bornmann et al., 2012). Cited references with higher impact would probably result in a higher impact of the citing document (Didegah and Thelwall, 2013).

The relationship between document features and citations has been studied extensively in the past. We have discussed the most relevant features for the process of citing.

### 3.1.2   Author features

Author features address aspects of a document beyond the document itself and its content. Authority influences the citing author's expectations of the document's value in general (Wang and Soergel, 1998). Well-known, and highly-cited authors (Bornmann and Daniel, 2007), authors well recognized by readers (Bjarnason and Sigfusdottir, 2002), and authors with higher academic rank (Pagel and Hudetz, 2011) attract more attention.

The organizational context of the document's author(s) affects the citing author's perception of the document's value. Several organizational features are reviewed by Tahamtan et al. (2016), including number of faculty members, number of published documents, number of productive and prestigious scholars in the department, rank of school or university, language of the institution, interdisciplinarity of the institutions, and the number of scholars in the organization. The nationality, race, age, and gender are also among the author's features by which the document value is assessed (Tahamtan et al., 2016). Documents written by authors from specific countries, with specific races, and specific gender are selected for citing more frequently than others (Tahamtan et al., 2016). For instance, documents published by authors from the US receive more attention than documents from other countries (Patterson and Harris, 2009).

The relation of citing author and cited document's author is influential in selecting and citing a document. A document that originates from a person or an organization which has a specific meaning for the citing author, such as friends, advisors, course professors, or former employers, is more likely to be read by the citing author (Wang and Soergel, 1998). It also happens that authors cite a document because its author is expected to be a referee of their developed product (Wang and White, 1999).

Co-authorships can be defined as the co-occurrence of two or more authors on a document (Frenken et al., 2010). International and national collaboration of authors lead to higher recognition of documents (Bornmann et al., 2014). One possible reason is that more authors present the results in scientific networks (such as conferences and workshops). Team diversity and interdisciplinary cooperation of co-authors (Antoniou et al., 2015), and the availability of former co-authors also increase a document's readership (and citedness) (Collet et al., 2014).

### 3.1.3 Journal features

The journal spectrum (centrality of the journal to the field) is one of the features by which documents are evaluated. Citing authors tend to look at documents in the few journals that are prominent in the scientific field. Thus, the authors assess a journal according to its standing in a field. The centrality of a journal to a field might mean that a weak document in a prestigious journal receives a considerable number of citations (Callaham et al., 2002). Furthermore, some multi-disciplinary journals (e.g. *Nature*, *Science*, or *Proceedings of the National Academy of Sciences of the United States of America*) attract cross-field attention; the same is true for journals from certain publishers. For instance, documents which are published in journals of the Nature publishing group receive high recognition among the scientific community (Vanclay, 2013).

The scope of a journal (specialized versus general and local versus international) results in its receiving more or less attention. Journals with international audiences (Annalingam et al., 2014), journals in fast-developing areas, such as genetics, biochemistry, and molecular biology (Huang et al., 2012), as well as English language (Bornmann et al., 2012) and interdisciplinary journals (Annalingam et al., 2014) are more likely to be read and cited.

### 3.1.4   Document values

Wang and Soergel (1998, p.121) define document values as "the user's perception of the desirability or potential utility of a document". Document values underlie the citing author's decision about whether or not to consider a particular document further (Wang and Soergel, 1998, Wang and White, 1999). Positive or negative attitudes toward a document's value in the citing author are related to the document, author, and journal features mentioned above. Five types of document values have been identified, including epistemic, functional, conditional, social, and emotional values. Documents' values are relevant in the selection, reading, and citation stages. Wang and White (1999) borrowed these values from the consumer theory of Sheth et al. (1991). They redefined the values that fit the context of information retrieval.

The epistemic value is defined as "the perceived utility of a document to satisfy a desire for knowledge or information that is unknown" (Wang and Soergel, 1998, p. 121). Sheth et al. (1991, p. 162) define the epistemic value as "the perceived utility acquired from an alternative's capacity to arouse curiosity, provide novelty, and/or satisfy a desire for knowledge". Some of the most important features according to which the epistemic value is understood are topicality, novelty, and discipline. Functional value is defined as the perceived utility of a document to contribute to the citing document. A citing author assesses a document's value as functional if it is topically relevant and therefore accepts it. The functional value can be influenced by features such as topicality, orientation/level, discipline, expected quality, and recency (Wang and Soergel, 1998).

In some circumstances, the perceived utility of a document is associated with social groups or persons, such as a famous researcher in the field, an advisor, colleague, organization, or other features of the authors. Then, the social context suggests that a document is valuable and should be cited (Wang and Soergel, 1998). Social value can be generally defined as the association with specific social groups, such as demographic, socioeconomic, or cultural-ethnic groups (Sheth et al., 1991). Authority mostly affects social value (Wang and Soergel, 1998).

Emotional value is defined as "the perceived utility of a document stemming from its capacity to arouse [positive or negative] feelings or affective states" (Wang and Soergel, 1998 , p. 122). So, if a citing author is positively emotionally affected by reading a document (e.g. because of its high quality or its publication in a reputable journal), he or she uses it more frequently than a document with a negative emotional effect (Wang and Soergel, 1998).

## 3.2   From selection to citation

Selection decisions of documents for citing are done by single citing authors, however other entities, such as author groups, institutions, or journals play significant roles in this process. They may have their own preferences in citing certain sets of documents. In the selection process, each document goes through two main processes: (1) Citing authors have different *reasons* to cite a document, e.g. the authors criticize the cited work. (2) Citing authors make a choice by applying *decision rules*.

### 3.2.1   Reasons to cite

Citing authors have a variety of reasons to cite documents. For instance, a citation can be set for historical reasons by tracking developments of the sciences. The same document can be cited by different authors following different reasons (Harwood, 2009). Also, the reasons for citing a particular document can vary over time (Case and Higgins, 2000).

In a first attempt, Garfield (1962, p. 85) notes a number of different reasons to cite: "paying homage to pioneers; giving credit for related work (homage to peers); identifying methodology, equipment, etc.; providing background reading; correcting one's own work; correcting the work of others; criticizing previous work; substantiating claims; alerting researchers to forthcoming work; providing leads to poorly disseminated, poorly indexed, or uncited work; authenticating data and classes of fact-physical constants, etc.; identifying original publications in which an idea was discussed; identifying the original publication describing an eponymic concept or term as, e.g., Hodgkin's disease, Pareto's Law, Friedel-Crafts Reaction, etc.; disclaiming work or ideas of others (negative claims); or disputing priority claims of others (negative homage)".

This first attempt already demonstrates that documents are cited for very different reasons, such as showing the theoretical connection between citing and cited documents, or documents are cited for reasons which are not related to research. The following list contains the main reasons for citing which have been proposed in the literature since the first attempt by Garfield (1962). Bornmann and Daniel (2008b) and Erikson and Erlandson (2014) review these reasons in more detail.

1) *Classical/founder:* If a document is the first substantial work on a topic, methodology, or technique or the document's author is the founder of a theory, method, or technique, the probability of citation increases. Citing authors tend to cite the first substantial work (Wang and White, 1999). This type of reasons for citation is deeply rooted in the normative theory of citing. Some studies found support for the existence of this type. The study by Riviera (2015, p. 1186) shows that "Italian scholars cite normatively, paying their intellectual debts and acknowledging colleagues' works". Similar conclusions have been published by Thornley et al. (2015) from an interview-based citation behaviour study. However, some classic and founder works seem to receive extraordinary recognition in a field.

2) *Persuasive*: The persuasive nature of citations is defined as the use of citations to establish knowledge claims, mostly by citing prestigious authors: authors cite to support their claims and to convince readers of their knowledge claims. Gilbert (1977) was among the first to provide such an interpretation of citations in his article entitled "referencing as persuasion". Persuasiveness has been found to be one of the most frequently referenced motivations (Brooks, 1985).

3) *Affirmational*: Chubin and Moitra (1975) introduced four types of affirmative citations: (i) if the findings of the citing document depend on the cited document, this is a basic essential citation. (ii) In subsidiary essential citations, the cited document is not directly connected to the subject of the citing document, but is still essential to the results of the cited document. (iii) Additional supplementary citation is the third citation type. Here, the cited document contains an independent supportive observation (idea or finding) with which the citing author agrees. (iv) The fourth type is the perfunctory supplementary citation, in which the document is cited without additional comments (Chubin and Moitra, 1975).

4) *Perfunctory*: Although Chubin and Moitra (1975) classify perfunctory citations as affirmative citations, Bornmann and Daniel (2008b) classify and describe them otherwise: citing authors makes a perfunctory reference to a cited document; the cited document is cited without additional comments; citing authors make a redundant reference to a cited document; and the cited document is not firmly relevant to the immediate concerns of the citing author. Perfunctory citations might be documents that are cited without being read (Latour, 1987).

5) *Critical or negational*: Authors cite a document in order to correct their own work or the work of others (Garfield, 1962). The work might be an example of a controversial work, disprove the results of the citing document, puts into question the data interpretation in the cited document, or disagree with an opinion or supposedly factual statement in the cited document. Two types of negational citations are mentioned by Chubin and Moitra (1975). *Partial negational citation:* the citing document disputes some aspects of the cited document and the citing author suggests that the cited document is erroneous in part and offers a correction. *Total negational citation:* the citing document negatively evaluates the cited document, corrects or questions it, or the citing author refers to the cited document as being completely wrong and offers alternative solutions. Vinkler (1987) suggests three levels of criticism: the cited document is criticized in minor points, the cited document is partly rejected or criticized in important points, or the cited document is fully rejected or completely criticized. The points of criticism may be: methodological weaknesses, lack of replication, insufficient reliability of the findings, or unsatisfactory interpretation of the findings. In a recent study, Catalini et al. (2015) analyzed citations from papers in the *Journal of Immunology* and identified 2.4% of the total number of citations as negative. The authors classified references as negative citations if they "pointed to the inability to replicate past results, disagreement, or inconsistencies with past results, theory, and literature" (Catalini et al., 2015, p. 13826).

6) *Methodological or operational*: Documents are cited because the citing authors point to used tools, practical techniques, materials, equipment, analysis methods, procedures, or study design (Moravcsik and Murugesan, 1975).

7) *Background reading* (historical background): Citations are used to trace intellectual influence and the history of knowledge claims. Thus, authors cite documents to point out further relevant information as part of the history of knowledge claims, or because they provide background reading and discussion.

8) *Understanding*: The author cites a document because it is essential for the understanding of the citing document.

9) *Conceptual:* Documents are cited because they present the concepts, definitions, interpretations, or theories which are used in the citing document. These are usually the original documents in which an idea, concept, or theory was introduced (Moravcsik and Murugesan, 1975). These citations can be called "concept markers" because they frequently represent particular concepts in the field (Case and Higgins, 2000).

10)  *Claiming*: Authors cite certain documents to defend their claims against attack. This situation is called positive claiming. For instance, (i) an author substantiates a statement or an assumption made in the citing document, or (ii) the results of the citing document proves, verifies, or substantiates the data or interpretation of the cited document (Spiegel-Rosing, 1977). It is also possible that the citing author disclaims the work or ideas of others or disputes priority claims of others (negative homage) (Garfield, 1962).

11)  *Alerting*: The citing author points to forthcoming works or refers the reader to additional reading (Garfield, 1962).

12)  *Contrastive:* Contrastive citations can usually be observed in the discussion section of documents. The author cites a document to contrast the results in the citing and cited documents. In the discussion section of citing documents, other works are contrasted with each other, or the results in the citing documents are presented as alternatives to the results in the cited documents (Bornmann and Daniel, 2008b). Moravcsik and Murugesan (1975) call these alternative citations juxtapositional citations.

13)  *Assumptive*: The citing author refers to assumed knowledge (that is a general or specific background), or the citing author acknowledges the work of pioneers (Bornmann and Daniel, 2008b).

3.2.2   Decision rules

Bazerman and Moore (2012) demonstrate that people usually take six steps when applying a rational decision-making. "(1) perfectly define the problem, (2) identify all criteria, (3) accurately weigh all of the criteria according to their preferences, (4) know all relevant alternatives, (5) accurately assess each alternative based on each criterion, and (6) accurately calculate and choose the alternative with the highest perceived value" (Bazerman and Moore, 2012, p. 3).

Decision-making is a search for good arguments. People wish to be able to justify their decisions that is, to have logical reasons why they made a choice. People also need a set of criteria for knowing when they are ready for decisions (Montgomery, 1983). Decision making typically takes place in a situation where enough information have been gathered that allows each potential choice to be evaluated and compared to the alternatives, and the decision-maker most often selects only one of them (Case and Given, 2016). In the decision making process, authors try to gather as much information and as many features as they can to reduce their feeling of uncertainty and to make their choice (Case and Given, 2016). Therefore, when the decision maker (e.g. the citing author) finds arguments that are strong enough for making a choice, it is proceeded by the feeling of confidence that the decision taken was the correct one to make (Montgomery, 1983, Vermeir et al., 2002). However, people vary in the level of confidence in their decisions made (Vermeir et al., 2002).

Citing authors have to make a choice regarding documents. They apply various decision rules, including *elimination*, *multiple-criteria*, *dominance*, *scarcity*, *sacrifice*, and *chain* to cite (or not) a certain document (Wang and Soergel, 1998, Wang and White, 1999). The *elimination rule* applies when the citing author considers only one criterion to evaluate the document values, which is sufficient to reject a document. In other words, a salient criterion is applied to reject a document without looking at other aspects of that document in decision-making (Wang and Soergel, 1998). For instance, a document is rejected for citing because it is published in a low-impact journal or describes a specific theory. Another document is rejected because it has not been published by an authority in the field. The *multiple-criteria rule* describes a situation where the citing author uses more than one criterion, such as the language of the document, its publisher, and the attractiveness of its topic to accept or reject a document (Wang and Soergel, 1998, Wang and White, 1999).

When several similar documents on a topic exist, citing authors tend to use the one that excels in at least one aspect, following the *dominance rule* (Wang and Soergel, 1998, Wang and White, 1999). For example, a book by the originator of a theory is cited instead of his or her paper because the book provides more details. Although a variety of decision rules have been proposed for how people choose among plenty of alternatives, it is asserted that the dominance rule is more fundamental than the others (Montgomery, 1983).

When only a few documents on a topic exist, the citing author is likely to apply less strict selection criteria, lower his or her threshold, and reads and cites less relevant documents. This is called the *scarcity rule* (Wang and Soergel, 1998, Wang and White, 1999). When – from the point of view of the citing author – enough documents on a topic are selected, the citing author does not read any more documents, and the selection and citation process on that topic is terminated. This is called the *satisfice rule* (Wang and Soergel, 1998, Wang and White, 1999). The last decision rule is the *chain rule*: citing authors prefer to make a collective decision on a set of documents that are connected to each other by a specific relationship, such as a document and its critiques or its citations. Another example is that the citing author makes a citation decision according to a special issue of a journal, so that a citation chain is applied (Wang and Soergel, 1998).

## 3.3   Citing document

Besides the features of the *cited document*, the features of the *citing document* also influence citation decisions. For example, the desire of co-authors to cite their own documents influences their selection of documents, or the target journal's instructions influence the authors' decision to cite certain documents. This section starts with empirical results on the location and number of citations (of the same document) in the citing document. Whereas the features described in section 3.1 focus on one document which is cited, citations (of the same document) can be included at different locations in the citing document.

### 3.3.1   Location and number of citations

Over the period from 1910 to 1983, the average number of cited references in documents was about ten. The number of references has trended upwards after this period (Swales, 1990) in all fields (Boyack et al., 2018). An important reason for this trend might be that citations were increasingly considered as central elements of academic writing (Bazerman, 1988, Hyland, 2004). In modern academic writing, citations maintain specific relevance to the citing document (author), and cited documents are discussed in more detail in all sections of the citing document (Swales, 1990, Hyland, 2004).

The location of citations in cited documents can be defined based on conventional paper sections (i.e. introduction, methods, results, discussion, and conclusions) (Swales, 1981), or even the beginning, middle, and end parts of the document. In general, citations are likely to be concentrated at the beginnings (e.g. introduction) and endings (e.g. discussion) of documents than in the middle (Boyack et al., 2018).

Citations are included in different parts of the citing document, chiefly based on the reasons and decision rules of the citing author (see section 3.2). Thus, the citation location might reveal the reasons and decision rules for citing a specific document. For instance, if a document is cited for conceptual reasons, it tends to be cited in the introduction (as part of the literature overview). Cano (1989) points out that citations in the introduction are more likely to present background literature than those in other sections. A citation in the discussion section is frequently about comparable results (Hu et al., 2015). A document might be cited more than once at different places in the citing document for different reasons (Carroll, 2016). Hu et al. (2015) show that for a repeatedly cited document, its first-time citation is usually perfunctory, unlike the later citations (see section 3.2.1).

Citations can have different relevance based on their locations in the citing documents (see section 3.1.4). Voos and Dagaev (1976) show that not only the reasons for citing a document, but also its relevance for the citing author can be deduced from its location in the citing document. The results of Safer and Tang (2009) in psychology reveal that the location predicts the importance of the cited reference. Bonzi (1982) used four categories to measure citation relevance: (1) not specifically mentioned in the text (e.g. "several studies have dealt with …"), (2) barely mentioned in the text (e.g. "Smith has studied the impact of …"), (3) one quotation or discussion of one point in the text (e.g. "Smith found that …"), and (4) two or more quotations or points discussed in the text. The scale that Bonzi (1982) introduces to measure citation relevance is based on the following premise: one measure of true relevance for a citing document is the extent of treatment of the cited in the citing document. A document which is cited without being discussed in the citing document can be expected to be less relevant than one which is discussed in-depth within the citing document (Bonzi, 1982). Boyack et al. (2018) note that there is a rough consensus in the literature that references located outside the introduction tend to be the most valuable.

Many studies have investigated the density of citations in different locations of a document. For instance, one study shows that the occurrences of citations are concentrated in the introduction (Hu et al., 2013). Another study indicates that the highest concentration of citations is in the first 15% of the citing document. According to this study, perfunctory citations encompass the largest category of citations in this section (33%). Organic citations comprise the largest category in the middle section (32%) and at the end (41%) of the cited document (Cano, 1989) (see section 3.2). Bornmann and Daniel (2008a) have studied the relationship between the location of citations within citing documents and the frequency of citations. Their study shows that 32% of the documents are cited in the introduction, 24% in the methods, 13% in the results, and 31% in the discussion sections. They indicate that documents with low or high citation counts are cited differently in the different sections of citing documents. For instance, documents with high citation counts are more frequently cited in the introduction (34% of the citing document) than documents with low citation counts (30% of the citing documents).

Bertin et al. (2016) analyzed the citation context of papers appearing in journals published by the Public Library of Science (PLOS). They classified the context whether the citing authors agree or disagree to the cited document. Their results show that "agreement is expressed mostly at the end of a research papers, especially in the 'Discussion' section and towards the end of the 'Results' section. Disagreement is less common in scientific discourse" (p. 1426). Boyack et al. (2018) indicate that the reference distributions vary in different fields of science (e.g. mathematics and computer science as well as social science and humanities).

The numbers of citations of specific documents and their location in the citing document depend also on the citing document features (e.g. the type of citing document), author features (e.g. the number of authors of the citing document), and journal features (e.g. guidelines of the journal).

### 3.3.2   Document features

Several document features, including document type and discipline of appearance, influence how documents are cited in the citing document. Generally, reviews cite documents in a different way than original articles (or notes and short communications). Reviews usually cite more documents than publications of another document type. Clarivate Analytics (formerly the IP & Science business of Thomson Reuters) even uses the number of cited references to classify documents as reviews or not. Any document containing more than 100 references and documents whose titles contain the word "review" or "overview" are coded as review (http://wokinfo.com/essays/impact-factor).

Besides document type, several studies have shown that the mean number of cited references varies with the field. Thus, the number of references in the citing document depends on its discipline. In Marx and Bornmann (2015), the average number of cited references in 2010 was higher for social sciences than for natural sciences, medical and health sciences, agricultural sciences, humanities, and engineering, respectively. Boyack et al. (2017) analyzed in-text citations of more than five million articles from PubMed Central Open Access Subset and Elsevier journals. They found similar reference distributions for biomedical and health sciences, life and earth sciences, as well as physical sciences and engineering: "citation counts of references peak at around the 30th centile, suggesting that methods papers are more highly cited than other types of papers" (p. 20). In contrast, mathematics and computer science as well as social science and humanities have more evenly distributed reference distributions. The number of self-citations in the citing document, which increases the total number of citations, also varies across different fields. Aksnes (2003) shows that the lowest percentage of self-citations in publications by Norwegian authors is in clinical medicine (17%) and the highest percentage is in chemistry and astrophysics (31%). Thus, discipline and topic of the cited document is related to the citations in the citing documents.

### 3.3.3 Author features

Features of the author (of the citing document) are central elements in the citation process. Authors are motivated to cite for scientific or non-scientific reasons (see section 3.2). Authors may cite appropriate sources properly and accurately. However, a work might also be cited after being transformed into something unrecognizable to the cited author (Cozzens, 1988).

The academic background of the citing author contextualizes the understanding, interpretation, and citation of relevant documents. Previous studies have shown that users' knowledge of a topic has significant effects on their search behavior for documents (Kelly, 2006). Users with high topic knowledge issue longer and more complex queries during their search than novice users (Hembrooke et al., 2005). Authors use their personal knowledge to interpret document values (see section 3.1.4), which are the basis for applying citation decision rules (see section 3.2.2). An author's personal knowledge can take a variety of types (Wang and Soergel, 1998), including topic knowledge (who are the relevant authors, when was the topic firstly discussed, and how is it related to the topic at hand), knowledge of the authors (who authored the document – knowledge which is often used to judge topical relevance), knowledge of the journal (by which the document's quality, orientation, and discipline are judged), and knowledge of the organization (which organization is associated with the document, which topics the organization works on, and how good its reputation is) (Wang and Soergel, 1998).

Besides academic background and individual attributes, we would like to mention three further author-related factors influencing citations: citation cartels, number of authors, and self-citations. Fister Jr et al. (2016, p. 1) define citation cartels as "groups of authors that cite each other disproportionately more than they do other groups of authors that work on the same subject". Thus, citation cartels explain relationships between citing and cited authors which are intended to increase the citedness of the citing authors' own papers (Fister Jr et al., 2016).

A higher number of co-authors on a citing document is usually related to a higher number of cited references. One possible reason is an increased number of self-citations. Glänzel and Thijs (2004) show that multi-authorship increases self-citation rates (and also the probability of being cited by others). Costas et al. (2010) and Aksnes (2003) similarly report that the number of self-citations increases with the number of authors of the citing document. The results of Aksnes (2003) and Glänzel and Thijs (2004) reveal, however, that the increase in the number of self-citations might play a minor role in the overall increase in citation rates. Although a higher number of citing authors can lead to more citations in the citing document, the number of self-citations can only explain a (small) part of the increased overall citation rates.

### 3.3.4 Journal features

The target journal where the authors are intending to submit their work might influence their decision about whether or not to cite a document (Wang and White, 1999). Some journals have their own expectations and demands for the references used. For example, they prefer some of their own documents to be cited in submitted manuscripts. Thus, authors consider the demands of journals to increase their chances of publication (Wilhite and Fong, 2012).

Some journals determine the number of references that should be used for different types of papers in their guidelines or instructions for authors. For instance, based on the "manuscript formatting guide" of *Nature*, articles (original reports) should have no more than 50 references. In *Science*, research articles should have about 40 references. Thus, due to such limitations or demands, documents relevant to the citing document might have no opportunity to be cited.

Citation cartels (see section 3.3.3) also refer to journal editors who use inter-journal citations to increase the JIF of their journals (Garfield, 2006). Thus, citation cartels additionally explain relationships between editor and citing author (Fister Jr et al., 2016) and thus the corresponding inclusion of cited references in citing documents.

## 4    Discussion

In this study, we provide an overview of the literature on the process of citing scholarly publications. The study is a conceptual overview which is structured according to three core elements, which concern the *cited* and *citing document* as well as the *process from selection to citation*. The overview starts with an explanation of three important citation theories, because many empirical studies in this area are rooted in these theories (especially the normative and constructivist theories). Authors have proposed further citation theories from various perspectives, such as the perspective of information retrieval (Cronin, 1984) or sociology (Cozzens, 1981). Leydesdorff (1998) described citations as something to be explained (explanandum) toward the articulation of a theory of citation. However, these further citation theories have not reached the importance of the normative and constructivist theories. Small (2004) and other researchers have pointed out that normative and constructivist theories should be combined in one theory (Liu, 1997). For instance, Cozzens (1989) presents a multi-dimensional model in which citations are part of the reward system of science, the rhetorical system of science, and the communication system of science. According to Leydesdorff (1998), however, a comprehensive theory of citation could not be formulated in the 20$^{th}$ century and even later on.

We would like to point to another theory of citation, which was proposed by Latour (1987) in his book entitled *Science in Action*. Luukkonen (1997) notes that this important theory has largely been ignored by the bibliometric community and discusses the reasons for such ignorance. According to this theory, the general role of citations is uniformly that of supporting knowledge claims. Authors use citations to support their knowledge claims and indicate with this that their work is scientific. Luukkonen (1997) notes that this explanation does not legitimate major uses of citation, its use as a performance measure, or as an indication of the development of specialties, as in co-citation analysis. Luukkonen (1997) further explains that one advantage of Latour's theory over the normative theory is that the former paints a coherent picture of the actual use of citations, and that citations perform different functions. Latour (1987) draws attention to the fact that citations vary widely from one situation to another and that authors have different reasons to cite. Luukkonen (1997) explains that these differences describe whether citations concern concepts or techniques or neither, whether they provide background reading, alert readers to an upcoming work, etc.

The conceptual overview of the literature in this study can be seen as an alternative to the proposed theories of citation, because it points to the most important elements – the core elements – in the citation process. These core elements have been identified in the manifold literature on the process and condensed in a schematic representation. The conceptual overview shows that citations in citing documents can be explained differently and depends on many factors. Besides simply counting the number of citations obtained, citations can be disaggregated based on their different values, including epistemic, functional, conditional, social, and emotional values (see section 3.1.4). Citations can also be disaggregated by the reasons of the citing authors (e.g. to support own knowledge claims or to criticize other research). These reasons can be assigned to different degrees of relevance for scientific progress: for instance, methodological or operational reasons can be assigned a higher relevance than perfunctory reasons. Thus, citations can simply be counted; however, using their perceived values in the citing document, new kinds of citation analyses seem possible (Boyack et al., 2017).

Authors consider a certain document further if they perceive some values in the document. These values determine the applied decision rules for selecting the document and reasons to cite. Each single citation process probably has its own characteristics across the core elements in the model and their assigned factors, rules, and reasons (Bornmann and Marx, 2012). For instance, some authors are included in a specific citation cartel and cite in a systematically biased manner. Other authors might find no value in a certain document, but cite it nevertheless (in a perfunctory way). The reasons for these citing decisions might arise from the requirements of the journal publishing the paper. The cartel might be organized around the journal, and the authors include citations of documents which are less relevant for the content of the citing paper, but have been published by the editors and potential reviewers of the journal. Another reason for citing specific documents might be the consensus of the scientific community on the importance of certain documents to the field which have become landmark papers or citation classics (Small, 2004). These citations might be added to documents in the field in a ritualized way which leads to an additional increase in citation rates of landmark papers or citation classics.

Besides document and journal features (on the cited and citing side), the authors' features (e.g. academic background and topic knowledge) influence the number of publications which are cited in a specific document and their locations in the citing document. For example, authors with a high level of topic knowledge find more information and more relevant information (Downing et al., 2005), while authors with incorrect or imprecise topic knowledge find less information and more irrelevant information (Keselman et al., 2008). The differing level of expertise in the field and (thus) the differing availability of publications from the literature searches influence what is cited as well as where and how frequently it is cited. For example, newcomers in the field do not know the original publications for certain methods, ideas, or important empirical results. Whereas experts in the field are able to cite these original publications, newcomers fail to cite these publications or cite later works which are based on and refer to these original publications.

The conceptual literature overview in this study can be helpful in different contexts of application of citation analysis. It can either be used to find answers on basic questions about the practice of citing documents: what possible reasons do the authors of a specific citing document have in making citation decisions? Are the reasons for citations in the document only related to the citing author or also to other people (e.g. the journal editor in which the document has appeared)? Do authors cite certain reputable scientists throughout the manuscript to increase the acceptance of their own line of research? Do the authors cite the relevant literature in a specific context (which point to high topic knowledge) or not (which point to low topic knowledge)? Which decision rules did the citing authors probably apply to select the cited documents from the literature? The overview can also be used as additional information for evaluative bibliometric reports (on single researchers, institutions, etc.): the reader of these reports are informed about the citation process for a proper interpretation of the bibliometric results. They learn, for instance, the long list of possible reasons to cite and the many factors which are associated with high or low citation counts (on the cited and citing documents' sides).

Besides answering basic questions on the process of citing (in general or in a specific case), the conceptual overview can also be used to obtain a better understanding of where scientometrics research on the citation process currently stands and where further research is still necessary. Our search of the literature revealed that many empirical studies focus on the cited document side (e.g. the relationship between the number of co-authors of the cited document and its citations), but only a few studies on the citing document side (e.g. the assumed requirements of journals for citing specific documents). One reason for this result of the overview might be that many citation studies are driven by the available data and full text is only recently becoming more easily available. Since, there is a lack of information regarding the relationship between features of citing documents and citations and how these features influence document selection and citation, future studies could focus on the following questions: How do journal features (e.g. members of the editorial board) influence the citation process? How do authors design their manuscripts to increase their chance of acceptance by certain journals (especially high-impact journals)? These questions can be answered by large-scale statistical analyses of bibliometric data on the citing side, in-depth analyses of citation context, and interviews with authors and journal editors.

For the empirical analysis of the citation process in future studies, new data sources can be used. The availability of citation context information in new sources of citation data, such as Microsoft Academic, opens new opportunities to study document values and possible reasons to cite (Haunschild et al., 2017, Hug et al., 2017, Hug and Brändle, in press, Zhang et al., 2013). Selection decisions can be assessed by empirically analyzing the online activity surrounding documents, including the number of times a document is viewed, downloaded, or saved from publishers' web sites and discussed or recommended on social media (Lin and Fenner, 2013). Counts of this data are called alternative metrics (altmetrics). One of the major providers collecting and analyzing altmetrics data is Altmetric (www.altmetric.com). The WoS and Scopus platforms also provide researchers with usage counts for publications, enabling researchers to analyze the usage data and to explore usage patterns (Wang et al., 2016, Thelwall et al., 2013).

# Acknowledgements

We would like to thank three anonymous reviewers and the chief editor of the *Journal of Informetrics* for valuable recommendations which lead to a substantially improvement of the manuscript.